# Locally Real States in the Elimination of Non-local Superposition and Entanglement

**Stuart Mirell[*] and Daniel Mirell**

*Quantum Wave Technologies, Los Angeles, CA, USA; https://quwt.com*

[*]*email: smirell@ucla.edu*

(Dated: July 4, 2023)

**ABSTRACT**

Locally real states of electromagnetic radiation derived from the underlying quantum mechanical formalism are shown to provide an alternative basis for definite polarized states of the widely accepted probabilistic interpretation. The locally real states traverse common loop configurations without invoking non-local superposition. Correlated photon pairs represented by the locally real states are consistent with the underlying quantum mechanical formalism independent of Bell's inequality and without invoking entanglement. Replacement of interdependent non-local probabilistic states with the locally real states yields independence of those states. A novel loop configuration provides differential testability of the representation of the locally real states relative to the probabilistic interpretation with the prediction of a unique polarization state.

**INTRODUCTION**

The underlying quantum mechanical formalism does not impose non-local phenomena. Rather, those phenomena are intrinsic to the widely accepted probabilistic interpretation of that formalism, PI. The non-local phenomena are attributed to particles and to photons in a variety of physical configurations as a necessary consequence of the positivist-based, compact PI.

An alternative to PI, a particular locally real representation, LR, derived from the underlying quantum mechanical formalism, provides locally real states for photons and for particles that do not exhibit non-local phenomena. [1]-[2] The LR representation, which inherently has the property of enhancement [3], is not subject to Bell's inequality. [4]

In the present report we focus on locally real states of photons and examine a variety of physical configurations in which the non-local phenomena of superposition and entanglement are absent. Among the most notable and frequently cited examples of non-locality are an individual photon in a "superposition" state transiting a loop configuration split into two separated amplitudes by polarization means and a pair of photons in an "entanglement" state, constituting two identical polarization states. The non-local basis for both of these examples can be traced back to the PI specification for a "definite polarized state."

For example, a photon transmitted by a polarizer is identified as being in a definite polarized state. That definite state is defined by the orientation of the polarizer's axis. If the axis orientation is designated 0°, the transmitted photon is specified as being in a definite 0°-polarized state. PI treats that photon as a non-real, probabilistic entity until a measurement is performed. That photon in a definite 0°-polarized state has a $\cos^2\theta$ probability of being transmitted upon encountering a polarizer rotated to some orientation θ.

Similarly, an atomic system that emits a correlated pair of PI-defined entangled photons is also inherently associated with definite polarized states. If a photon of a pair is transmitted by the polarizer of one Bell detector, that photon is specified, by virtue of the detector measurement, of probabilistically realizing a particular definite polarized state upon interacting with that polarizer. With the polarizer's axis orientation defined as 0°, that state is more precisely identified as a definite 0°-polarized state. Accordingly, following that photon's measurement by the first Bell detector, the other photon of the pair, in transit to the other Bell detector's polarizer, is then also in a definite 0°-polarized state from the perspective of PI, a perspective that imposes the non-local property of entanglement on the correlated pair.

As an alternative to PI, we examine the consequences of treating photons as objectively real entities in the LR representation of states. This treatment assigns to any single photon an objectively real wave structure. Deterministically, that wave structure prescribes the transmissibility of the photon through an encountered polarizer. These assignments have important consequences in the context of photons in definite polarized states.

In this report we specifically compare the PI and LR treatments of photons in definite polarization states for a variety of configuration cases. In those cases for which PI necessarily invokes non-locality, we show that the LR treatment is unambiguously and self-consistently local. In other cases, the LR treatment elucidates physical processes that are occluded in the PI treatment.

Before examining these cases, below we very briefly review LR with particular regard to structural aspects of photon wave structure and interaction of that structure with polarizers. These topics are covered in great detail in ref.'s [1] and [2].

LR is effectively definable as a locally real "hidden variable theory," HVT. The "hidden variables" are seen to be the transverse wave packet structure derived from first principles of the underlying quantum mechanical formalism that "completes" the three-dimensional representation of an objectively real wave packet. [5] LR, which inherently includes the property of enhancement, is not subject to Bell's theorem, but is consistent with reported results of Bell experiments and PI predictions of those experiments. [1]

Longitudinally, the objectively real wave structure of a photon is consistent with that given by the wave function of the underlying quantum mechanical formalism. Transversely, that wave structure radially spans an arc Δ. For an "ordinary" full complement photon, that arc span is 90°. A notable exception includes atomic emission correlated pairs of photons where, for partial complement photons, that arc span may be less than 90°. A radial arc bisector defines the orientation of a particular photon. Fig. 1 illustrates a two-wavelength segment of an ordinary linearly polarized, full complement photon wave packet oriented at -40° (and equivalently at +140°). The energy quantum associated with the photon resides at a point on the wave structure with a locus probability proportional to the squared modulus of the wave amplitude.

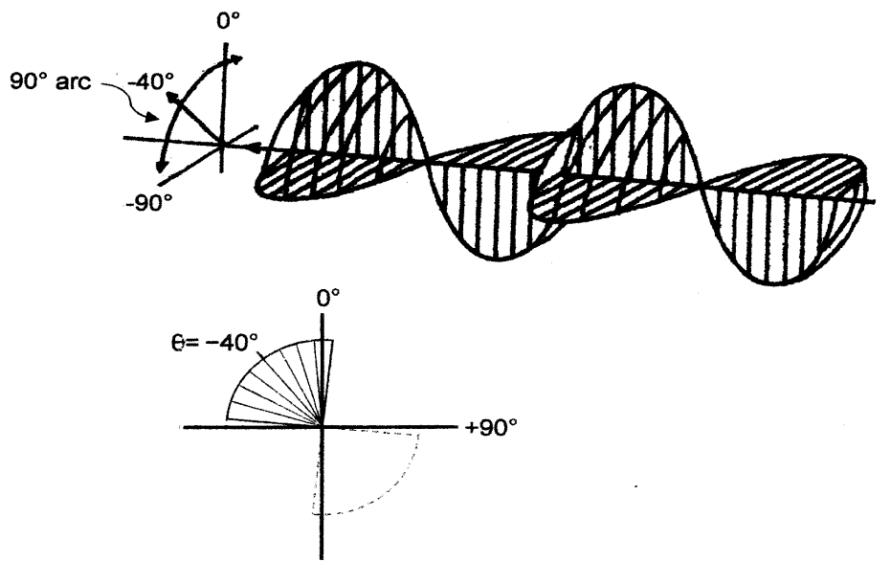

*Fig. 1. A projective view of a two-wavelength long segment of a linearly polarized photon wave packet with an orientation (arc bisector) at -40° as detailed in the accompanying transverse cross-sectional view.*

An objectively real elliptically polarized wave packet in LR is a physically local superposition of two transversely orthogonal waves of the form associated with objectively real, linearly polarized photons.

The respective amplitudes and the relative phase of the two wave forms determine the interaction properties of an elliptically polarized wave packet.

From the perspective of LR, the PI representation of a photon is incomplete. The inclusion of a transverse structure derived from the quantum formalism together with the longitudinal wave structure represents an objectively real state which, when applied to a broad range of quantum configurations, eliminates non-locality.

We begin with the LR photon wave packet's interaction with polarizers. If the Fig. 1 depicted 3-D wave packet encounters a two-channel polarizer, the wave structure is split. If the polarizer's axes are at 0° and 90°, axial projection generates two objectively real planar 2-D waves propagating respectively along the 0° and 90° axial planes within the polarizer media. In that projective process the relative amplitude magnitudes of the two planar waves on the respective axes are cos 40° and cos 50°.

Additionally, as the 3-D wave structure enters the polarizer, that structure condenses as the planar waves begin to form. Since the 0° axis intersects the 3-D structure's arc span, the condensation process effectively "traps" energy quantum on the forming 0° planar wave, then appropriately designated as an "occupied" wave. The 90° planar wave, absent the energy quantum, forms as an "empty" wave.

If the polarizer is instead one-channel, only one planar wave is formed within the polarizer media. For example, if the polarizer's single axis is at 0°, only the same reduced amplitude 0° planar occupied wave forms in the polarizer. Conversely, if the polarizer's single axis is at 90°, only the same reduced amplitude 90° planar empty wave forms in the polarizer and the energy quantum, no longer sustained by a wave, is absorbed by the polarizer media.

The representation [1] treats "observables", i.e. the objectively real occupied waves on which the energy quanta reside. Nevertheless, the empty waves are also objectively real entities and in the interests of completeness, those empty waves are treated in the fully inclusive representation of LR considered as in ref. [2].

Complementary to the above interaction of LR objectively real states with polarizers, is the LR polarization ensemble that statistically represents the objectively real states of a definite polarized state. When an incident photon is successfully transmitted in a polarizer that ensemble is generated at the exit face of the polarizer and a random member of the ensemble is emitted. The polarization ensemble consists of an infinite-member set, each with a 90° arc span, and with orientations (arc bisectors) defined by a cosine squared envelope centered about the orientation of the definite polarization state which for an emitting polarizer is the axis of the polarizer.

In most of the discussions below, "ensemble" refers to the ubiquitous "polarization ensemble" unless otherwise specified. The notable exception considered below is the "correlated photon ensemble" that relates to LR states of photons defined by PI as "entangled."

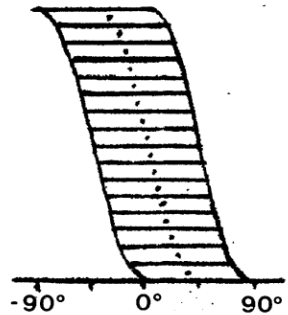

*Fig. 2. An infinite-member polarization ensemble is closely approximated by the depicted 16-member ensemble. For that depicted 0°-polarized ensemble, a random member (row) statistically represents the orientation (arc bisector) of a photon emitted from a polarizer oriented at 0°.*

For instructional purposes the actual infinite-member ensemble can be closely approximated by the finite member ensemble depicted in Fig. 2. If the polarization axis of the polarizer is 0°, the depicted 0°-centered ensemble is the appropriately oriented LR finite-member approximation of a PI definite 0°-polarization state. A random state (row) of that ensemble constitutes an objectively real emitted member of the ensemble. The rows represent the arc spans of the individual ensemble members with their respective orientations (arc bisectors) indicated by dots.

The equation

$$\theta°_\alpha = \cos^{-1}[(\alpha-1)/(N-1)]^{1/2} - (\theta+45°) \tag{1}$$

is the general representation of a finite-member ensemble, giving the bisector orientations for a finite number N of ensemble members where the ensemble centered at the θ orientation. In Eq. (1) the finite-member ensemble is represented by the quantity $\theta_\alpha$ where θ is the orientation of the polarization ensemble and the values of α, 1→N identify the individual members with α=1 corresponding to the lowest row.

For the particular Fig. 2 ensemble where θ=0° and N=16, Eq. (1) takes the form

$0°_α = \cos^{-1}[(α-1)/(15)]^{1/2} - 45°$.  (2)

The finite-member polarization ensembles represented by Eq.'s (1) and (2), are functionally identifiable as "oriented-member ensembles."

It can readily be appreciated that a vertical axis applied to the Fig. 2 distribution at some angular orientation θ transects a $\cos^2θ$ fraction of the arc spans. With that vertical axis physically representing the polarization axis of a one-channel polarizer intercepting the arc spans of random members of the ensemble, we see that a well-defined $\cos^2θ$ fraction of the members are deterministically transmitted by the polarizer.

In this representation of the oriented-member ensembles, the arc bisectors of a finite set are mapped one-to-one onto the precise orientation associated with each of the respective members. In the limit as N→∞, the oriented-member ensemble converges to an infinite-member ensemble that precisely represents a physical polarization ensemble.

An instructive functional alternative to the above "oriented-member ensemble" is constructed from a uniform angular partitioning of space. The resultant "partition-based ensemble" is generated from the derivative of the arc-bisector curve given by

$B(θ) = \cos^2(θ+45°)$  (3)

for a physically precise infinite-member 0°-polarization ensemble. Eq. (3) is recognized as the right-hand envelope boundary of the ensemble translated to the left by 45° thereby bisecting each of the N-infinite members. The derivative of B(θ),

$∂_θ B(θ) = -\sin[2(θ+45°)]$  (4)

is proportional to $ΔN/Δθ_i$ which is a function of $θ_i$. The $Δθ_i$ are uniform angular partitions of space applied to the transverse plane, each of magnitude δ and centered at a $θ=θ_i$. ΔN is the fraction of N members associated with a particular $Δθ_i$ partition This relationship accurately holds for small δ together with finite but suitably large N giving a finite-member ensemble that trivially converges to a precise representation of an actual physical ensemble as δ→0°.

Conversely, as δ increases to macroscopic angular partitions, exact proportionality of $ΔN/Δθ_i$ to $∂_θ B(θ)$ degrades. Nevertheless, at a very coarse δ=30° the resultant partition-based ensemble is remarkably useful. A mere four members provide a complete set for the δ=30° partition ensemble that retains functional equivalence to the corresponding actual physical ensemble at the $θ_i$'s of the partition ensemble. Additionally, because of these functional equivalences, the four-member, δ=30° partition-based ensemble is symbolically utilized below in representation of actual physical ensembles.

Fig. 3 depicts that four-member, partition-based, 0°-polarization ensemble identified as the set of "emission" photons $\{γ_e\}$. The four members are individually identified as A, B1, B2, and C to facilitate tracking of each state in the process of various ensemble interactions.

The "generator" photon $γ_g$ in Fig. 3 identifies the planar wave photon in the polarizer that gives rise to the ensemble $\{γ_e\}$ at the exit face of the polarizer from which a random member is emitted. Notably, the generator photon $γ_g$, transversely occupying a finite 30° partition in the Fig. 3 partition-based ensemble, properly converges to its actual physical planar form in the limit as δ→0°. (The designation of {$γ_e$} is equally applicable to the Fig. 2 oriented-member ensemble.)

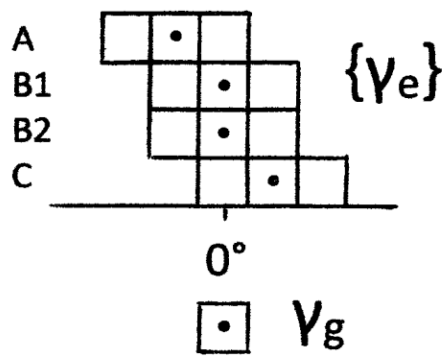

*Fig. 3. A four-member, partition-based, (emission) ensemble {γ_e} symbolically and functionally representing a 0°-polarization ensemble. (This partition-based, finite-member ensemble is substantially a representational alternative to the Fig. 2 oriented-member, finite-member ensemble.) This figure additionally includes the planar state γ_g from which {γ_e} is generated.*

The transmission of each individual member of the Fig. 3 ensemble is fully deterministic with respect a subsequent polarizer oriented to the partition orientations ±90°, ±60°, ±30° and 0°. For the entire four-member ensemble, a polarizer oriented to the complete set of partition orientations -90°, -60°, -30°, 0°, +30°, +60°, and +90° yields respective transmission factors of 0, 0.25, 0.75, 1, 0.75, 0.25, and 0, after normalization by a factor of four, in exact agreement with transmission factors for a definite polarization state.

The Fig. 2 finite-member oriented-member ensemble and the Fig. 3 partition-based ensemble both converge to an infinite-member ensemble in their respective limits, N→∞ and δ→0°. Both finite-member ensemble constructions are useful in the analyses below concerning the relationship of PI definite polarization states and LR locally real states. These constructions are also relevant to the analysis below of PI correlated states in relation to LR locally real states.

**CASE STUDIES IMPLEMENTING LOCALLY REAL STATES OF PHOTONS**

We analyze the implementation of LR locally real photon states in cases relating to a broad range of physical configurations. The respective analysis results are compared to those in which the PI probabilistic photon states are implemented.

**Case 1-Correlated Photons associated with an atomic emission process**

The first case examined here concerns correlated photons and differs from the subsequently examined cases which all explicitly relate to definite states of polarized photons. The PI treatment of correlated photons as non-locally entangled entities is widely regarded as the principal basis for rejecting local realism. Accordingly, a comparative analysis of the respective PI and LR treatments of correlated photons is essential to the present work.

We begin with the analysis of correlated photons produced in an atomic transition when measured by a Bell experiment. [6]-[7] When a photon $γ_1$ of an emitted correlated pair $γ_1$ and $γ_2$ is detected after being transmitted by a Bell detector polarizer oriented at $θ_1≡0°$, PI contends, based upon that measurement process, that $γ_1$ was in a definite 0° polarization state. Because of the quantum correlation relationship of $γ_1$ and $γ_2$ arising from their emission process, PI implicitly requires that the measurement puts $γ_2$ in the same definite polarization state as that of $γ_1$. Accordingly, the transmission outcome of $γ_2$ upon encountering the other Bell polarizer oriented at some selected θ is statistically predictable as proportional to $cos^2θ$. This prediction is consistent with $γ_2$ being in the definite 0° polarization state.

Each transmission event of a $γ_1$ photon conducted in the photon reference frame is equivalent to a related experiment in the laboratory reference frame of a single photon emitter directing its photons to a first polarizer oriented at 0° followed by a second polarizer with its axis polarization at θ. In this experiment, any photon transmitted by the first polarizer is in a definite 0° polarization state and the resultant transmission outcome at the second polarizer is necessarily $cos^2θ$.

From this PI perspective we see the $γ_2$ photon merely as a photon in a common definite polarization state with that of $γ_1$. The common definite polarization state can be statistically confirmed by choosing a variety of θ settings for the second polarizer relative to the 0° orientation of the definite state. However, the cost of invoking that common polarization state for $γ_2$ is that of relinquishing locality. The definite state of $γ_2$ is dependent on the setting of the remote second Bell polarizer θ. For ordinary photons such as $γ_1$ and $γ_2$, that dependency requires non-local entanglement of the photon pair since local correlation of the pair at their emission site cannot take into account the setting of the remote θ.

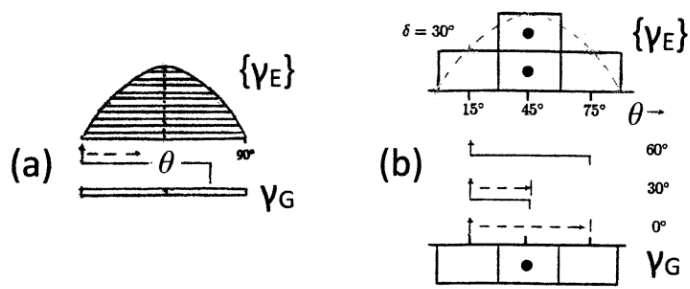

*Fig's. 4a-b. (a) Locally real states of correlated photon pairs measured by Bell detectors. The pair consists of a generator photon $γ_G$ and a random member of a correlated photon emission ensemble $\{γ_E\}$, all members of which have an orientation in common with that of the generator photon. Statistically, with the Bell detector polarizers orientations separated by θ, joint transmission is proportional to $cos^2θ$. (b) A partition-based two-member ensemble very succinctly demonstrates that joint transmission result. For both figures, the ensemble envelope is given by sin 2θ.*

LR, which is not subject to Bell's theorem because of the inherent property of enhancement, [3] demonstrates that one of the pair mates, a "generator" photon $γ_G$, is a full complement 90° arc span photon. From first principles of the underlying quantum mechanical formalism it is shown that $γ_G$ generates a correlated photon "emission" ensemble $\{γ_E\}$ from which a random member is emitted as $γ_E$. [1-2] The $γ_E$ are in general asymmetric to the $γ_G$ in the regard that the $γ_E$ have a distribution spectrum of arc spans rather than the fixed 90° arc span of $γ_G$. Unlike the $γ_e$ of a polarization ensemble, the $γ_E$ are all mutually degenerate in orientation within a sine envelope. For example, when the orientation of $γ_G$ is defined as 45° that {γ_E} ensemble is defined by a sin 2θ envelope. In that example, Fig. 4a depicts the correlated photon ensemble for a large number of members in a close approximation to the actual infinite member ensemble. The LR representation of the correlated photon's γ_G generator state and the associated correlated photon emission ensemble {γ_E} is dissimilar to that of the polarizer generator photon γ_g and its associated polarization emission ensemble {γ_e}. Nevertheless, when joint-detection is evaluated for γ_G and the associated correlated emission state ensemble {γ_E}, the $cos^2θ$ relationship normally associated with polarization states naturally emerges as a consequence of local correlation at the time the pair of correlated photons is still confined to the atom.

That $cos^2θ$ relationship is most expediently demonstrated by applying the Fig. 4b coarse finite 2-member ensemble for the emission photon ensemble {γ_E} in which a 30° angular partition is applied. Over that partition, sampling ratios of 4/4, 3/4, 1/4 and 0/4 are acquired when the Bell polarizer orientations are relatively rotated by 0°, 30°, 60° and 90°, giving agreement with $cos^2θ$. As δ→0° and the number of ensemble members increases toward ∞ as in Fig. 4a, that agreement with $cos^2θ$ extends in a continuum over all θ representing relative angular rotations of the two Bell polarizers. In LR that agreement simply arises from local correlation at the time of atomic emission.

Conversely, that agreement is treated by PI as confirmation that a detection of one probabilistic photon of a pair at a first Bell detector instantly entangles the probabilistic pair mate into a definite polarization state in a non-local process. The perception of entanglement is a consequence of the subtlety of enhancement. [1]

**Case 2**. **State replication in a fully deterministic locally correlated loop.**

We turn next to a fully deterministic configuration, a loop comprised of two opposed, contiguous calcite crystals illustrated in Fig. 5. This case is examined in [2] in detail that is not repeated here. However, from that work we can immediately deduce the output results when a vertically (0°) polarized state is input to the loop. For calculational expediency we can represent that vertically polarized input state with the Fig. 6 vector diagram which functionally and symbolically compactly represents the Fig. 3 partition-based ensemble of a vertically (0°) polarized state.

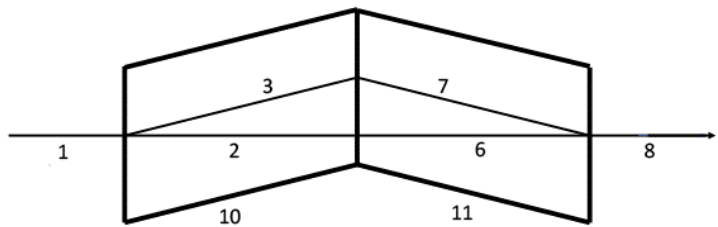

*Fig. 5. A loop comprised of contiguous calcite crystals. An input photon expressed as a locally real state is deterministically replicated at the loop output without non-local superposition as a result of local correlation that occurs at the loop input.*

Because the loop is completely deterministic, the state of each objectively real input state at any θ, where -45°<θ<+45°, is exactly replicated at output. Accordingly, symbolically Fig. 6 is a vector diagram of the loop input states as well as the loop output states. For example, the "A" input state of the Fig. 6 LR polarization ensemble at -30° is replicated as a -30° output state. This process continues for each input state in the continuum limit as δ→0°, trivially replicating the entire input set of states at the loop output. Since the input set in the present case happens to have the characteristic distribution of a polarization ensemble, the output distribution under one-for-one replication will trivially also have that characteristic distribution. Moreover, under one-for-one replication, any distribution would be replicated because of the local correlation that occurs as each photon enters the loop. No non-local superposition is present.

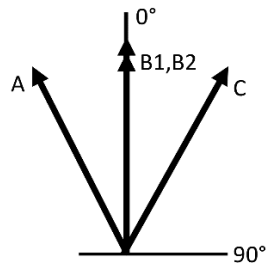

*Fig. 6. A symbolic vector diagram of a vertical (0°) polarization ensemble depicting the four locally real states of the partition-based Fig. 3 ensemble $\{\gamma_e\}$.*

In contrast, PI treats the definite 0°-polarized input state as a probabilistic entity that does not admit any objectively real structure to that input state. That treatment necessarily imposes non-local superposition in the loop transit process.

Ultimately, the LR and the PI treatments both yield statistically indistinguishable output results but with the PI treatment necessarily imposing non-local superposition.

**Case 3**. **State replication in a loop with local correlation that includes random orientation.**

An example of this case is a PBS MZ loop shown in Fig. 7 along with a detail of the first PBS dielectric layer beam paths. This case is also examined in [2] at length and is not repeated here. However, there are important aspects of that case that were not stressed in that reference, namely how local correlation still results in one-for-one replication despite the random process that occurs at the output of the first PBS.

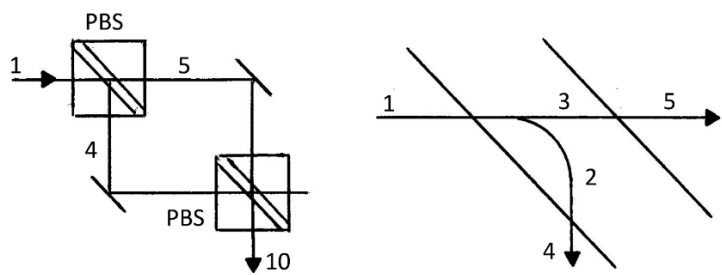

*Fig. 7. A PBS MZ loop in which the output state deterministically replicates the input state without non-local superposition despite a random orientation that occurs as split states emerge from the first PBS dielectric layer shown in the figure detail. Those split states are shown to be locally correlated at that first PBS.*

At that first PBS, the two outputs each yield polarization ensembles from which one random member is resultant. One output ensemble is a vertical polarization ensemble that can be symbolically represented by the Fig. 6 vector diagram and the other output ensemble is a horizontal polarization ensemble, basically the Fig. 6 set of vector states rotated clockwise by 90°, as shown in Fig. 8. However, the resultant random members of those two output ensembles are correlated to each other. Under time reversal symmetry, the two output members must have orthogonal orientations in order to equal the input state. That means, for example, that if it is member C of the vertical ensemble that is emitted then it is member C of the horizontal ensemble that is the accompanying emitted member. This local correlation is all that is needed to ensure that LR yields a final output state from the second PBS that replicates the objectively real input state and also imposes a 90° rotation on that state.

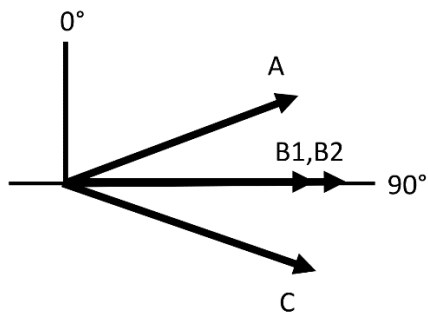

*Fig. 8. A symbolic vector diagram of a horizontal (90°) polarization ensemble depicting the four locally real states of the partition-based Fig. 3 ensemble {γ$_e$} rotated by 90°.*

Consequently, as in Case 2, this process continues for each input state trivially replicating the entire input set of states at the loop output. Since the input set happens to have the characteristic distribution of a polarization ensemble, the output distribution under one-for-one replication will trivially also have that characteristic distribution. Symbolically, a Fig. 6 input would result in a Fig. 8 output. As in Case 2, this process continues for each input state in the continuum limit as δ→0°, trivially replicating the entire input set of states at the loop output. Moreover, under one-for-one replication, any distribution would be replicated with 90° rotation because of the local correlation that occurs as each photon enters the loop. No non-local superposition is present.

As in Case 2, the LR and the PI treatments both yield statistically indistinguishable output results with the PI treatment necessarily imposing non-local superposition.

**Case 4**. **Replication of a polarization state without individual state replication**

This case presents an interesting contrast to Cases 2 and 3. Case 4 relates to reproducing a conventional definite PI polarization state at the output of a loop such as the dis-joint opposed calcite polarizers in Fig. 9 despite the absence of local correlation and superposition.

For instructional purposes in this analysis, without loss of generality, we consider a diagonal definite polarization state input on path 1 of Fig. 9. We begin with a detailed examination of this loop case since it was not examined earlier in [2] unlike the loop cases 2 and 3.

In LR the diagonally polarized input state is represented by $\Phi_1(45°_\alpha)_+$ where α specifies some random member of the $45°_\alpha$ diagonally oriented polarization ensemble. A large number of such input photons would constitute the equivalent of the photons associated with a PI diagonal definite polarization state.

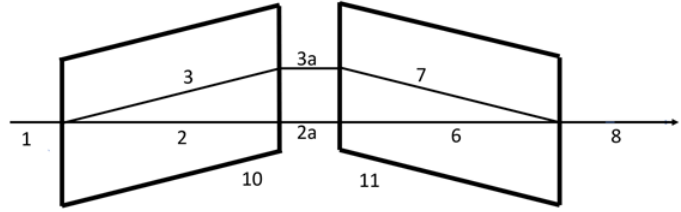

*Fig. 9. A loop comprised of non-contiguous calcite crystals. An input photon expressed as a locally real state is statistically replicated at the loop output without non-local superposition as a result of the global symmetry properties of the loop.*

The LR state

$$\Phi_1(45°_\alpha)_+ = \xi\, \mathbf{r}(45°_\alpha)_+ \tag{5}$$

and $\xi$ represents the normalized longitudinal wave function. The radial unit vector **r** is oriented at the arc bisector of an objectively real α member of the 45°$_\alpha$ diagonally oriented polarization ensemble. [2]

When $\Phi_1(45°_\alpha)_+$ is incident on calcite 10 (the first calcite), the state is split into the two planar wave states

$$\Phi_{\delta 2}(0°)_+ = \xi \cos(45°_\alpha)\, \mathbf{r}_\delta(0°)_+ \tag{6}$$

and

$$\Phi_{\delta 3}(90°) = \xi \sin(45°_\alpha)\, \mathbf{r}_\delta(90°). \tag{7}$$

In this representation of the two planar wave states, Eq.'s (7) and (8), we have again imposed an initial condition on the orientation of the incident photon for instructional purposes and without loss of generality. In this condition the random ensemble member α is associated with a member that has its arc bisector orientation θ confined to 0°<θ<+45° (as opposed to +45°<θ<+90°). As a result, the energy quantum residing on $\Phi_1(45°_\alpha)_+$, is transferred to the vertically projected state component $\Phi_{\delta 2}(0°)_+$ as indicated by the appended "+" subscript that is not appended to the horizontally projected state component $\Phi_{\delta 3}(90°)$. This imposed condition allows us to progressively track the energy quantum as the wave states transit the loop but does not otherwise alter the wave states. $\Phi_{\delta 2}(0°)_+$ is then identifiable as a photon or, equivalently, an occupied wave whereas $\Phi_{\delta 3}(90°)$ is an empty wave. (Distinct states are numerically identified by paths on which properties such as transverse arc spans, wave amplitude and energy quantum occupation are invariant.)

At the exit face of calcite 10, the physically separated planar waves $\Phi_{\delta 2}(0°)_+$ and $\Phi_{\delta 3}(90°)$ respectively emit the states

$$\Phi_{2a}(0°_\mu)_+ = \xi \cos(45°_\alpha)\, \mathbf{r}(0°_\mu)_+ \tag{8}$$

and

$$\Phi_{3a}(90°_\nu) = \xi \sin(45°_\alpha)\, \mathbf{r}(90°_\nu). \tag{9}$$

These two states are very similar to the states $\Phi_4(0°_\alpha)_+$ and $\Phi_5(90°_\alpha)$ emitted from a polarizing beam splitter PBS 25 regarding Figs. 5 and 6 from ref. [2] but with an important distinction. Because of their physical separation, the μ and ν ensemble members emitted from the calcite 10 in Fig. 9 are mutually uncorrelated with respect to arc bisector orientation. In contrast, because of conservation constraints, the conjoint planar states $\Phi_{\delta 2}(0°)_+$ and $\Phi_{\delta 3}(90°)$ within the PBS 25 dielectric emit $\Phi_4(0°_\alpha)_+$ and $\Phi_5(90°_\alpha)$ that have arc bisector orientation correlation, i.e. the two states are random members of polarization ensembles but have mutual α-correlation with respect to those ensembles.

When the uncorrelated $\Phi_{2a}(0°_\mu)_+$ and $\Phi_{3a}(90°_\nu)$ are incident on calcite 11 of Fig. 9, they respectively generate planar wave states

$$\Phi_{\delta 6}(0°)_+ = \xi \cos(45°_\alpha) \cos(0°_\mu) \, \mathbf{r}_\delta(0°)_+ \tag{10}$$

and

$$\Phi_{\delta 7}(90°) = \xi \sin(45°_\alpha) \sin(90°_\nu) \, \mathbf{r}_\delta(90°), \tag{11}$$

that propagate in calcite 11. At the exit face of calcite 11 those planar wave states vectorially combine to produce

$$\Phi_{\delta 6}(0°)_+ + \Phi_{\delta 7}(90°) = \xi \, [\cos(45°_\alpha) \cos(0°_\mu) \, \mathbf{r}(0°) + \sin(45°_\alpha) \sin(90°_\nu) \, \mathbf{r}(90°)]_+ = \Phi_8(45°_\alpha)_+ \tag{12}$$

In the those particular instances when μ=ν, cos (0°$_\mu$)=sin (90°$_\nu$) and Eq. (12) reduces to the compact form

$$\Phi_{\delta 6}(0°)_+ + \Phi_{\delta 7}(90°) = \cos(0°_\mu) \, \xi \, \mathbf{r}(45°_\alpha) = \Phi_8(45°_\alpha)_+ = \cos(0°_\mu) \, \Phi_1(45°_\alpha)_+. \tag{13}$$

For the particular instances in which μ=ν the output state $\Phi_8(45°_\alpha)_+$ would replicate the orientation of the input state $\Phi_1(45°_\alpha)_+$ aside from having the amplitude reduced by a projection factor such as cos (0°$_\mu$). (Recall that for Case 2, a conjoint calcite loop, and for Case 3, a PBS MZ loop, the orientation of the output state replicates the orientation of each loop input state. Then, trivially, an input PI definite polarization state is replicated at the loop output for Cases 2 and 3.)

However, in the presently considered of the Case 4 disjoint calcite loop, μ and ν are uncorrelated. Generally μ≠ν for the disjoint calcite loop and the less compact form of $\Phi_8(45°_\alpha)_+$ Eq. (12) must be used to evaluate the output state orientation.

That evaluation can be conducted by a Monte Carlo simulation. In that simulation, for every α value of an input polarization ensemble, all possible values μ and ν must be evaluated in all possible combinations. The simulation is most succinctly and instructively conducted for an N=4 member polarization partition-based ensemble similar to that of Fig. 3 where angular increments are still partitioned into 30° intervals but where the {γ$_e$} ensemble is centered at 45° instead of 0° in order to properly represent a diagonally polarized ensemble.

Fig. 10 depicts the corresponding vector representation of that diagonally polarized ensemble with member states at 15°, 45°, 45° and 75°. The α member indices 1, 2, 3, and 4 may be assigned respectively to the vector states C, B2, B1, and A. The Fig. 6 vertically polarized ensemble with member states at -30°, 0°, 0° and +30° and the Fig. 8 horizontally polarized ensemble with member states at +60°, +90°, +90° and +120° are also similarly used in the simulation respectively for quantities having μ and ν indices. The simulation involves $4^3$=64 evaluations of Eq. (12).

The simulation shows that for each α input state the orientation of the output state from the loop averages to the orientations 15°, 45°, 45° and 75° associated with the four μ=ν combinations. From the perspective of LR for the disjoint calcite loop of Case 4, an input state orientation does not precisely map to the same orientation on output, other than for those four where µ=v, however the simulation shows that statistically each LR input state maps to a functionally equivalent output state at the same orientation as depicted in Fig. 11. This equivalence constitutes a statistical variant of the one-for-one mapping of cases 2 and 3. Then, statistically, all members of an input polarization ensemble are functionally replicated at the output. More generally, under one-for-one replication, any input distribution is functionally replicated at the output.

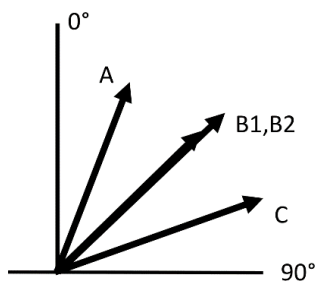

*Fig. 10. A symbolic vector diagram of a diagonal (45°) polarization ensemble depicting the four locally real states of the partition-based Fig. 3 ensemble {γ$_e$} rotated by 45°.*

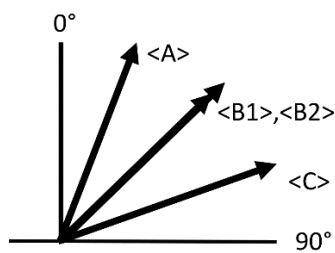

*Fig. 11. A symbolic vector diagram depicting the four locally real averaged states emerging from the Fig. 9 loop when the Fig. 10 states are input.*

Importantly, in the limit as the angular partition δ→0° and the number N of ensemble members approaches ∞, the LR distribution of output states is indistinguishable from that of PI. This case is profound because it shows that it is the global symmetry of the Fig. 9 dis-joint calcite loop analyzed with the LR states that results in functional replication of the input ensemble without the local correlations of Cases 2 and 3 and without invoking superposition as in PI.

**Cases 5-7**. We depart from considerations of superposition in the following but continue here in demonstrating that the inclusion of LR objectively real states reveals underlying objectively real physical properties that are not representable by PI.

**Case 5**. **State transmission through a half wave plate, HWP.**

The half wave plate, HWP, is a totally deterministic device related to the Fig. 4 contiguous opposed calcites in that regard. In PI, 0° (vertically) polarized photons traversing a HWP with its fast axis at 45° rotates the photons to horizontally polarized photons. (There are two propagation paths as in loops but these paths are colinear and they do not relate to nonlocal superposition phenomena.) LR however shows that the Fig. 6 states map to the Fig. 12 states which is indeed a 90° rotated polarization ensemble but the ensemble has also undergone a mirror reflection. A subsequent HWP at 90° corrects that mirror reflection and results in a purely rotated ensemble, Fig. 8. The point here is that the inclusion of LR states reveals objectively real structure that is obfuscated by PI.

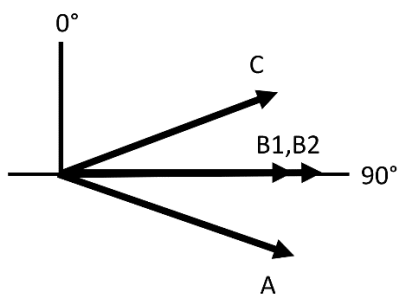

*Fig. 12. A symbolic vector diagram of a horizontal (90°) polarization ensemble depicting the four locally real states of the partition-based Fig. 3 ensemble {γ$_e$} rotated by 90° and reflected about the horizontal (90°) axis.*

**Case 6**. **Malus's Law.**

In ref. [1] Malus's Law is shown to be a trivial demonstration of LR in which photons successfully traversing a conventional linear polarizer as planar wave photons have a "lost history" of their prior orientation. (The transmitted photons do objectively incur a characteristic loss of wave intensity related to their prior orientation but that parameter is not of immediate relevance to Malus's Law.) With each emission from a polarizer, the emitted photon is a random member of a θ-polarization ensemble where θ is the polarization axis of the polarizer. The inclusion of LR states shows that the outcome of any polarizer-emitted photon having a particular objectively real state is fully deterministic with regard to transmission or absorption in a subsequent polarizer of defined orientation.

**Case 7**. **Random polarization states and conjoined vertical and horizontal polarization states.**

PI defines a polarization state by a measurement process such as rotation of an imposed polarizer that demonstrates a cos$^2$θ transmission probability. We can demonstrate a trivial example where there is a distribution of objectively real states fully representable in LR but for which PI is necessarily oblivious. Moreover, the representation of a randomly polarized state as a superposition with basis vectors at any orientation is shown to be in inconsistent for states that statistically lack rotational symmetry.

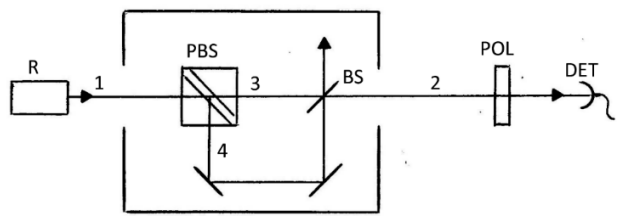

*Fig. 13. A thought experiment in which detected photons appear to be randomly polarized despite objectively being either in a definite vertical or horizontal polarization state.*

Fig. 13 shows a source R emitting a beam of single randomly polarized photons on beam path 1 that enters a box. The beam exits the box on path 2 and is directed to a rotatable polarizer followed by a detector.

Photon detection is observed to be independent of the polarizer orientation which is consistent with the random polarization of the source R.

However, from the perspective of LR and unseen by the PI experimenter, inside the box each photon arriving on path 1 is directed by a PBS onto path 3 or path 4 depending on whether the photon was oriented +45°<θ<+135° or -45°<θ<+45°, respectively. In that process, the photon on path 4 is then symbolically a random member of an ensemble such as {$\gamma_e$} in Fig. 3 and the photon on path 3 is symbolically a random member of the Fig. 3 ensemble but centered at 90°. Because of the two mirrors and conventional 50:50 beam splitter also in the box, the photons from paths 3 and 4 are combined onto path 2. The individual photons are well separated and non-interacting as they proceed to the rotatable polarizer and detector.

Fig. 14 shows a representative statistical distribution of the individual photons on beam path 2. Note that because the horizontal and vertical ensembles "nest," the number of states intersected by the polarizer's axis and are transmitted is independent of the polarizer's orientation θ. The detector would appear to show that the photons are randomly polarized whereas from an LR perspective the orientations (arc bisectors) of the objectively real photons are characteristically clustered about 0° and 90°.

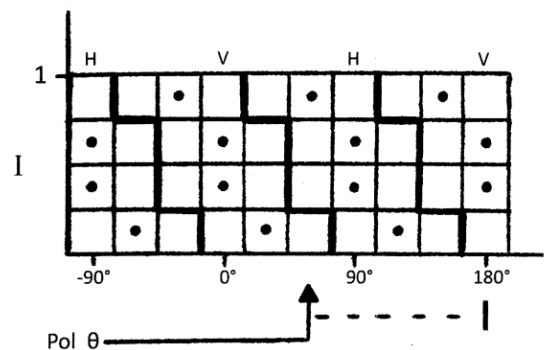

*Fig. 14. Measured irradiance I of the Fig. 13 experiment normalized to unity represented on partition-based ensembles. Each detected photon is statistically assigned to one of the ensemble member states intersected by the polarizer's axis at θ. At θ=60° 25% of those photons are objectively members of a vertically polarized ensemble. At θ=0° 100% of those photons are objectively members of a vertically polarized ensemble. The experimental result of θ-invariant irradiance does not distinguish the vertically and horizontally polarized photons from randomly polarized photons.*

This thought experiment demonstrates another example of LR showing a microstructure, the orientation distribution of constituent photons, that is not evident from the associated measurement process. PI judges the photon beam on path 2 as randomly polarized and therefore functionally representable as a (local) superposition of horizontally and vertically polarized photons independent of the orientation of the basis vectors.

While that independence is functionally correct, LR shows us that basis vectors specifically at 0° and 90° correspond to orientations of actual physically real polarization ensembles. The clustering of state bisector orientations about 0° and 90° cannot be measured and PI is consistent in treating the photons on path 2 as randomly polarized. While this deficiency does not rise to level of imposing a non-local phenomenon, it does represent a further omission of objectively real structure from the perspective of LR.

**Cases 8 and 9. Configurations for testable differentiation of PI and LR.**

In the configurations associated with the foregoing 7 cases the LR representation self-consistently predicts properties of resultant states that are objectively real but for which the resultant states themselves are not testably distinguishable from the corresponding resultant states predicted by PI. The question arises, are there configurations for which a self-consistent application of the LR representation predicts objectively real properties of resultant states that render the resultant states testably distinguishable from those of the corresponding PI predicted states? This question is addressed with the Case 8 and Case 9 analyses of two closely related configurations depicted respectively in Fig.'s 15 and 16. These configurations differ by their respective output components, a two-channel polarizer such as calcite and a polarizing beam splitter but the configurations are functionally equivalent.

Fig.'s 15 and 16 depict a loop receiving vertically polarized photons on path 1 from source S. The photons are input to a conventional 50:50 beam splitter. For both configurations, the optical path lengths of the loop legs are equal.

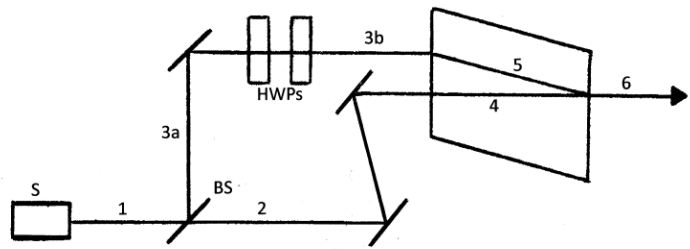

*Fig. 15. A loop configuration for splitting the amplitude of a vertically polarized photon into two equal magnitude amplitudes and rotating one of those amplitudes by 90° whereby the respective equal but orthogonal vertical and horizontal axial projections of the two amplitudes vectorially combine in a two-channel polarizer such as calcite to produce a super polarized photon oriented at 45°.*

We first examine the states from the perspective of PI for the Fig. 15 and 16 configurations. The beam splitter divides the input state $|v\rangle_1$ on path 1 into $(\sqrt{2}/2)|v\rangle_2$ on path 2 and $(\sqrt{2}/2)|v\rangle_{3a}$ on path 3a. Two consecutive HWP's on path 3a, the first at 45° and the second at 90°, rotate the vertically polarized path 3a state to the horizontally polarized state on path 3b, $(\sqrt{2}/2)|h\rangle_{3b}$.

For the Fig. 15 configuration the path 2 state $(\sqrt{2}/2)|v\rangle_2$ is directed to the calcite polarizer where it propagates along path 4 as $(\sqrt{2}/2)|v\rangle_4$ to the exit face of the polarizer. The path 3b state $(\sqrt{2}/2)|h\rangle_{3b}$ is directed to the calcite polarizer where it propagates as $(\sqrt{2}/2)|v\rangle_5$ to the exit face of the polarizer in spatial and temporal coincidence with $(\sqrt{2}/2)|v\rangle_4$. At the exit face of the polarizer the states on paths 4 and 5 combine to yield a path 6 output state

$$(\sqrt{2}/2)|v\rangle_4 + (\sqrt{2}/2)|h\rangle_5 = |d\rangle_6. \tag{14}$$

For the Fig. 16 configuration, as shown in the PBS dielectric layer detail, the path 2 state $(\sqrt{2}/2)|v\rangle_2$ is directed to the PBS dielectric layer and the path 3b state $(\sqrt{2}/2)|h\rangle_{3b}$ is directed to the same point on the dielectric layer in spatial and temporal coincidence with $(\sqrt{2}/2)|v\rangle_2$. In the dielectric layer of the PBS the states on paths 2 and 3b combine to yield a path 6 output state

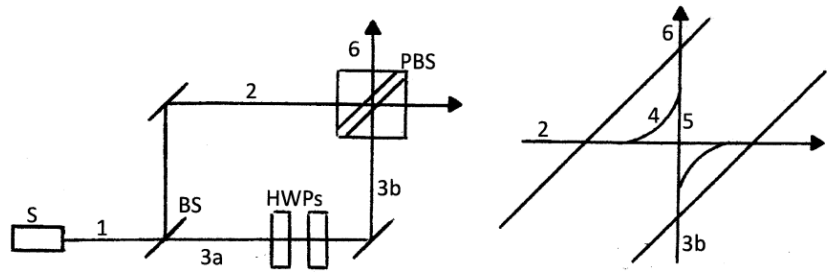

*Fig. 16. A loop configuration closely analogous to that of Fig. 15 whereby the respective equal but orthogonal vertical and horizontal axial projections of the two amplitudes vectorially combine in a polarizing beam splitter to produce a super polarized photon oriented at 45°.*

$(\sqrt{2}/2)|v\rangle_2 + (\sqrt{2}/2)|h\rangle_{3b} = |d\rangle_6.$  (15)

In both cases PI unremarkably predicts a definite 45°-polarization state for photons on path 6. A rotatable polarizer intercepting the path 6 photons should yield a characteristic cosine squared transmission probability distribution centered at 45°.

The PI predictions for the path 6 output states can be symbolically represented by the Fig. 17 45°-polarization ensemble and the associated Fig. 10 vector diagram.

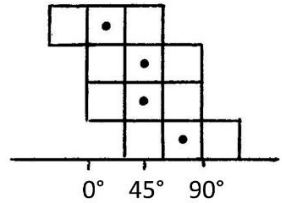

*Fig. 17. A symbolic partition-based representation of a 45°-polarization ensemble.*

The corresponding LR analysis of the Fig. 15 and 16 configurations is conducted by considering an objectively real member of the input vertical polarization ensemble. That member has an orientation (arc bisector) at some θ where -45°<θ<+45°. We track the outcome of that single objectively real member. On path 1

$\Phi_1(\theta)_+ = \xi\, \mathbf{r}(\theta)_+$  (16)

where **r** is a unit radial vector in the transverse plane oriented at θ. The path 2, 3a, and 3b states are respectively $\Phi_2(\theta) = \xi\,(\sqrt{2}/2)\,\mathbf{r}(\theta)$, $\Phi_{3a}(\theta) = \xi\,(\sqrt{2}/2)\,\mathbf{r}(\theta)$, and $\Phi_{3b}(\theta+90°) = \xi\,(\sqrt{2}/2)\,\mathbf{r}(\theta+90°)$.

We shall shortly see that the entire loop transit process is independent of the particular loop leg randomly taken by the energy quantum of the input photon. Accordingly, on the states intermediate to the input state $\Phi_1$ and the output state $\Phi_6$ the occupation specification "+" is omitted.

For Fig. 15 when states $\Phi_2$ and $\Phi_{3b}$ are incident on the calcite polarizer, those states proceed in the polarizer on paths 4 and 5 as the planar wave states

$\Phi_{64}(0°) = \xi\,(\sqrt{2}/2)\cos\theta\,\mathbf{r}_6(0°)$  (17)

and

$\Phi_{\delta 5}(90°) = \xi\,(\sqrt{2}/2) \sin(\theta+90°)\,\mathbf{r}_\delta(90°)$

$\qquad = \xi\,(\sqrt{2}/2) \cos(\theta)\,\mathbf{r}_\delta(90°).$  (18)

As $\Phi_{\delta 4}$ and $\Phi_{\delta 5}$ emerge from the exit face of the calcite polarizer they vectorially add to

$\Phi_{\delta 4} + \Phi_{\delta 5} = \xi\,(\sqrt{2}/2) \cos\theta\,\mathbf{r}_\delta(0°) + \xi\,(\sqrt{2}/2) \cos(\theta)\,\mathbf{r}_\delta(90°)$

$\qquad = \xi\,(\sqrt{2}/2) \cos\theta\,[\mathbf{r}_\delta(0°) + \mathbf{r}_\delta(90°)]$

$\qquad = \xi\,(\sqrt{2}/2) \cos\theta\,\sqrt{2}\,\mathbf{r}(45°)_+$

$\qquad = \xi \cos\theta\,\mathbf{r}(45°)_+$

$\qquad = \Phi_6(45°)_+$  (19)

For Fig. 16 when states $\Phi_2$ and $\Phi_{3b}$ are incident on the PBS dielectric layer, those states proceed in the dielectric layer of the PBS on paths 4 and 5 as the planar wave states

$\Phi_{\delta 4}(0°) = \xi\,(\sqrt{2}/2) \cos\theta\,\mathbf{r}_\delta(0°)$  (20)

and

$\Phi_{\delta 5}(90°) = \xi\,(\sqrt{2}/2) \sin(\theta+90°)\,\mathbf{r}_\delta(90°)$

$\qquad = \xi\,(\sqrt{2}/2) \cos(\theta)\,\mathbf{r}_\delta(90°).$  (21)

which vectorially add as they emerge from the PBS dielectric layer giving

$\Phi_{\delta 4} + \Phi_{\delta 5} = \xi \cos\theta\,\mathbf{r}(45°)_+$

$\qquad = \Phi_6(45°)_+.$  (22)

The output states in Fig. 15 and 16 configurations are functionally identical. Both cases show that an arbitrarily oriented input member of a vertically polarized ensemble is mapped to a 45° orientation. $\Phi_6(45°)_+ = \xi \cos\theta\,\mathbf{r}(45°)_+$ is a single objectively real photon generated on the output path 6 of the Fig. 15 and 16 configurations from a single objectively real input photon state $\Phi_1(\theta)_+ = \xi\,\mathbf{r}(\theta)_+$ that has an orientation $\theta$ where $-45° < \theta < +45°$. A subsequent input photon state $\Phi_1(\theta')_+ = \xi\,\mathbf{r}(\theta')_+$ that has an orientation $\theta'$ generally distinct from $\theta$, but still subject to $-45° < \theta' < +45°$, similarly results in an output state $\Phi_6(45°)_+ = \xi \cos\theta'\,\mathbf{r}(45°)_+$. This is a fully deterministic outcome consistent with the same principles employed in other fully deterministic outcomes such as those associated with the Fig. 5 and Fig. 7 loops.

All of the subsequent input photons result in output photons that have identical orientations of 45° and differ only by the cosine amplitude coefficient that they each acquire as objectively real members of the vertically polarized ensemble of input photons. Those coefficients range from 0.707 to 1 and do not differentially distinguish the various output photons on path 6 with regard to measurement by a rotatable polarizer and a detector. For a sampling of four input photons, Fig. 18 is a symbolic representation of the distribution of photon states output onto path 6 from the LR analysis. Fig. 19 depicts the corresponding vector representation of those states.

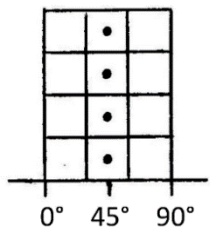

*Fig. 18. A symbolic partition-based representation of an ensemble of super polarized photons oriented at 45°. Four members are depicted but since all members are identically oriented, a single photon would constitute a representative ensemble. A clockwise rotating polarizer oriented at θ receiving a beam of these 45°-oriented super polarized photons is predicted to yield step function full transmission beginning at 0° continuing uniformly to a complementary step function non-transmission at 90°.*

The PI-based Fig. 17 ensemble distribution is distinguishable from the LR-based Fig. 18 ensemble distribution with a rotatable polarizer and a detector. The Fig. 17 distribution represents common diagonally (45°) polarized photons. Fig. 18 represents objectively real "super polarized" photons at 45°.

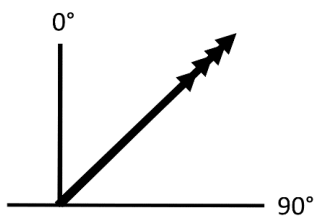

*Fig. 19. The symbolic vector diagram representing the Fig. 18 super polarized ensemble.*

It can be appreciated that the essence of the above process resolves to splitting a photon wave oriented at some θ into two waves of equal magnitude still oriented at that θ and then rotating one of those waves by 90°. When those waves each respectively encounter a channel in a two-channel polarizer, the respective vector projections along the vertical axis and horizontal axis of the polarizer are equal and add to an output vector at 45° independent of the orientation of the input photon. Super polarizations to orientations other than 45° can readily be achieved by using a BS with a transmission:reflection ratio other than 50:50.

The super polarization process is dependent upon the LR objectively real structures of photons that are inherently inadmissible and not representable in PI.

Recall that for the Fig. 4 and Fig. 6 loops, LR shows that input states are replicated one-for-one at the outputs. In the above we have self-consistently applied the same principles used in those two cases to show how any objectively real input state at some θ associated with an input vertically polarized photon, can be deconstructed and reconstructed to give a state at 45°.

Note that in the above Fig. 15 and 16 configurations the second HWP, which is oriented at 90°, could be omitted. This omission would not result in the 90° rotation of the state that had been on path 3a but would instead result in the mirror image of the 90° rotated state on path 3b about the horizontal axis of the PBS. The resultant projection coefficient $\cos(-θ)$ is mathematically the same as the $\cos θ$ coefficient that occurs with the inclusion of the second HWP. In any case, the inclusion of the second HWP is preferred in the interest of generating an objectively real orthogonal rotation of LR states on path 3a.

**DISCUSSION**

In this report we have deduced particular locally real states consistent with LR that are commonly associated with the ubiquitous PI definite polarization states. From the LR analysis of correlated photons, distinctively different locally real states have also been deduced. Historically, the PI treatment of correlated photon states has profoundly impacted the treatment of all quantum phenomena. We further examine that PI treatment of correlated photons here in part because of that impact and in part because the PI treatment improperly conflates the definite polarization states and the correlated photon states from the perspective of LR.

For a correlated photon pair, $γ_1$ and $γ_2$, PI postulates that when $γ_1$ is successfully measured by transmission through the 1$^{st}$ Bell polarizer oriented at 0°, the $γ_2$ mate is instantly in a definite correlated state relative to that of $γ_1$ constrained by angular momentum conservation. In the positivist-based PI that definite correlated state is implicitly equivalent to a definite polarization state that should be transmitted by the 2$^{nd}$ Bell polarizer with the characteristic cosine squared distribution associated with polarizer emission.

In PI the definite polarization state is a ubiquitous entity that is generated most notably by a photon emitted from a polarizer. That state is experimentally demonstrated by analysis with a rotatable polarizer that statistically yields the classic cosine squared transmission distribution.

For correlated photons, from the perspective of LR, that $γ_2$ mate of the correlated pair is not in a state generated from a definite polarization state. There is no polarizer present from which a definite polarization state would be generated. LR shows that it is the objectively real planar state photon propagating within a polarizer and transversely confined at the orientation of the polarizer axis that is responsible for generating the ensemble associated with a definite polarization state. That planar state is absent in the correlated pair. LR self-consistently shows that it is the correlated pair photon with a full complement 90° arc span and designated as the generator photon $γ_G$ ($γ_1$), that generates an "emission ensemble" {$γ_E$} from which a random member $γ_E$ ($γ_2$) is emitted without incurring non-local entanglement.

The necessity of invoking non-locality in PI can be understood from the associated positivist principles that reject the inclusion of objectively real but non-measurable structures in the representation of quantum states.

LR shows that it is incorrect for PI to interpret that $γ_2$ of the correlated pair is in the same "definite polarization state" as $γ_G$ ($γ_1$). $γ_E$ ($γ_2$) must conserve angular momentum relative to $γ_G$ because the pair arises in free space. This conservation is satisfied in LR by the associated "emission ensemble" {$γ_E$} (that should more precisely be named "correlated photon emission ensemble"). The emission ensemble is shown to be generated from the full 90° arc span of the "generator" photon. All members of that emission ensemble have the same orientation, ensuring conservation of angular momentum with respect to the generator photon. Despite the differences in the definite polarization ensemble {$γ_e$} and the correlated photon emission {$γ_E$} with its spectrum of arc spans, we still get a cosine squared transmission distribution with the polarizer of the 2$^{nd}$ Bell detector. Moreover, since the mutual alignment of the generator photon $γ_G$ and the {$γ_E$} ensemble occurs locally at the atomic system, the correlation is local.

In contrast polarizer emitted states do not conserve angular momentum relative to the planar photon that generates those states because the polarizer emission process includes the polarizer itself. The output states of the polarizer exhibit a spectrum of orientations relative to the planar generator photon.

That spectrum is incompatible with the constraint of angular momentum conservation for all members of an appropriate correlated photon emission ensemble. However, that constraint is satisfied by the LR $\{\gamma_E\}$ in which all members have the same orientation.

Analyses of correlated photons with the $\{\gamma_e\}$ definite polarization state ensemble, as is implicit in PI, or with the $\{\gamma_E\}$ correlated photon emission ensemble, as in LR, yield the same cosine squared transmission distribution albeit with non-locality in the former. In that regard, the analyses provide still another example in which the absence of objectively real non-measurable structures in PI obfuscates potentially underlying, distinguishing processes which in the present case are consistent with a local reality.

In a related context the implementation of the LR locally real states for correlated photons, as well as more generally, for photons in definite polarized states results in the elimination of the "measurement problem." The origin of that problem is seen to reside with the non-local phenomena of PI and not with the underlying quantum mechanical formalism.

**CONCLUSION**

From the underlying quantum mechanical formalism, a self-consistent, objectively real structure is derived. For photons this structure treats the usual longitudinal wave function as an objectively real structure that is "completed" to a three-dimensional structure by the inclusion of the objectively real transverse structure derived from that formalism. [1], [2], and [5] The completed structure constitutes the basis for locally real states in a particular representation of local reality, LR, that is inherently not subject to Bell's inequality. [1], [2], and [3]

The LR locally real states are applied here in the analyses of photon phenomena associated with a variety of commonly encountered physical configurations. Phenomena necessarily associated with non-locality by the probabilistic interpretation, PI, are represented by the LR locally real states without invoking non-locality. In the analyses of these cases of commonly encountered physical configurations, the LR predicted measurable results, while fully in agreement with those of PI, do not provide the means to differentially distinguish between PI and LR.

The impasse of distinguishing between PI and LR is addressed here. Two closely related, novel physical configurations are deduced for which the objectively real structure of LR does intervene to predict testable results that differentially distinguish between PI and LR. For these two configurations, with input photons in a definite polarization state, PI predicts that the output photons are unremarkably also in a definite polarization state. Conversely, a self-consistent implementation of the LR locally real states predicts that all of the output photons are "super polarized" with identical orientations.